\documentclass[
superscriptaddress,
groupedaddress,
reprint,
nofootinbib,
amsmath,amssymb,
aps,
prd,
floatfix,
]{revtex4-2}
\usepackage[utf8]{inputenc}
\usepackage{graphicx}
\usepackage{dcolumn}
\usepackage{bm}
\usepackage{hyperref}
\hypersetup{
    colorlinks=true,
    allcolors=blue,
}

\usepackage{xcolor}

\usepackage{tikz}
\usetikzlibrary{decorations.pathmorphing}
\usepackage{mathrsfs}

\begin{document}

\title{Revisit Static Aether: Exact Vacuum Solution in Einstein-Aether Theory and Its Analytic Extension}
\author{Jie Zhu}
 \email{jiezhu@cqu.edu.cn}
 \affiliation{Department of Physics and Chongqing Key Laboratory for Strongly Coupled Physics, Chongqing University, Chongqing 401331, P.R. China}

\author{Hao Li}
  \email{Corresponding author: haolee@cqu.edu.cn}
   \affiliation{Department of Physics and Chongqing Key Laboratory for Strongly Coupled Physics, Chongqing University, Chongqing 401331, P.R. China}

\date{\today}

\begin{abstract}
We obtain an exact analytical vacuum solution of Einstein-Aether theory with a strictly static aether configuration and investigate its maximal extension. The solution depends only on the coupling $c_{14}$ and reduces to the Schwarzschild geometry in the limit $c_{14}=0$. We show that Schwarzschild is an isolated member of this family: for any nonzero $c_{14}$ the spacetime ceases to be a black hole and instead becomes either a naked singularity ($c_{14}<0$) or a wormhole-like geometry ($0<c_{14}<2$). By constructing the complete analytic extension, we demonstrate that the internal infinity of the wormhole corresponds to an extremal Killing horizon. Crossing this horizon leads to a new spacetime region where the causal roles of time and radial coordinates are exchanged, and the spacetime ultimately terminates at a spacelike singularity. The resulting global structure, summarized by the corresponding Carter-Penrose diagrams, reveals a previously unexplored causal completion of the static-aether vacuum spacetime.
\end{abstract}

\maketitle

\section{Introduction}
General Relativity~(GR) and the Standard Model~(SM) of particle physics are the most successful theories describing all four fundamental forces of nature.
However, there are theoretical tensions between GR and SM, and to reconcile them, several candidates of quantum gravity~(QG) theories have already been proposed.
Generally, the onset of significant effects of QG is expected to happen at the Planck scale~(\(E_{Pl}\sim 10^{19}~\text{GeV}\)), which is far beyond our reach for current experiments.
Although direct detection of QG effects seems to be unlikely at present, it is suggested that there exists the possibility that certain kinds of remnant signals of QG could be observed at energy scales much lower than the Planck scale.
One of such signals is the violation of Lorentz invariance.
To explore potential signatures of Lorentz invariance violation within the gravitational sector, various alternative gravity frameworks have been proposed; prominent among these are Einstein-aether (EA) theory~\cite{Jacobson:2000xp, Eling:2003rd, Jacobson:2004ts, Eling:2004dk, Foster:2005dk}, Bumblebee gravity~\cite{Kostelecky:2003fs, Bluhm:2004ep}, and Hořava gravity~\cite{Horava:2009uw, Mukohyama:2010xz, Wang:2017brl}.

In this work, we specifically restrict our interest to EA theory. As a prominent Lorentz-violating framework, it extends GR by introducing a dynamical, unit-timelike vector field $u^a$ known as the aether. This privileged field singles out a preferred local rest frame at each point in the manifold, thus providing a natural and self-consistent testing ground for exploring quantum gravity phenomenology in the low-energy regime.
The first spherical static vacuum solutions in the EA theory were obtained by Eling and Jacobson in 2006~\cite{Eling:2006df}.
Spherically symmetric vacuum spacetimes in EA theory have been studied extensively in the past couple of years, both analytically~\cite{Berglund:2012bu, Berglund:2012fk, Gao:2013im, Ding:2015kba, Ding:2016wcf, Lin:2017cmn, Ding:2018whp, Oost:2019amf, Azreg-Ainou:2020bfl, Chan:2019mdn, Chan:2020amr, Churilova:2020bql, Khodadi:2020gns, Rayimbaev:2020gqj, Oost:2021tqi, Chan:2021ela}, and numerically~\cite{Eling:2006ec, Eling:2007xh, Tamaki:2007kz, Blas:2011ni, Barausse:2011pu, Zhu:2019ura}.

Although numerous vacuum solutions have been found in EA theory, major theoretical challenges remain. 
A prominent puzzle is the vacuum solution featuring a strictly static aether configuration, where $u^a$ is restricted to a vanishing spatial component.
This static configuration was initially investigated by Eling and Jacobson in 2006 as a one-parameter family depending entirely on the coupling $c_{14} = c_1 + c_4$~\cite{Eling:2006df}. They found that for $0 < c_{14} < 2$, the asymptotically flat solutions manifest a wormhole topology with an internal infinity that acts as both a curvature singularity and an extremal Killing horizon. However, due to an inconvenient choice of coordinate systems, the explicit and concise exact analytical expressions remained unavailable in their work.
Subsequently, Chan et al.~\cite{Chan:2020amr} revisited the static aether configuration. Hindered by an unsuitable coordinate choice, they only obtained vacuum solutions for several discrete parameters ($c_{14} = -16, 0, 3/2, 16/9, 48/25, 2$) in highly complicated expressions, leaving the general exact solution for arbitrary $c_{14}$ still unaddressed.
More recently, Oost et al.~\cite{Oost:2021tqi} obtained the static aether vacuum solutions in isotropic coordinates; however, a comprehensive discussion regarding the global causal and physical properties of this spacetime is still lacking.
In particular, for this class of solutions, it remains unknown whether a black hole configuration forms when $c_{14} < 0$, whether the internal infinity acts as a Killing horizon for $c_{14} > 0$, and what the nature of the spacetime is beyond this horizon.

In this paper, we show that the static aether vacuum solutions possess a remarkably simple analytical form under a well-chosen coordinate patch. Through analytic extension, we obtain a previously undiscovered branch where the signs of $g_{tt}$ and $g_{rr}$ flip. We verify that the solutions represent naked singularities when $c_{14} < 0$. 
Within the parameter space $0 < c_{14} < 2$, by analytically extending or smoothly connecting the two regions, the extended geometry enables test particles to cross the internal infinity into a completely new spacetime region, eventually hitting a final spacelike singularity. A detailed analysis of the global causal structure for various $c_{14}$ regimes is also provided.

The structure of this paper is outlined as follows. Section~\ref{sec:EA} introduces EA theory and its equations of motion. Section~\ref{sec:sol} presents the explicit derivation of the vacuum solutions. The geometric classification of these solutions under various parameter regimes is carried out in Sec.~\ref{sec:geo}. Section~\ref{sec:PSGC} is devoted to the analysis of photon spheres and geodesic completeness. In Sec.~\ref{sec:Analytic}, we construct the analytic extensions and investigate the complete global spacetime structure. Section~\ref{sec:summary} concludes the paper with a brief summary.

\section{Field equations in the EA theory}\label{sec:EA}
The general action of the EA theory is given by
\begin{equation}
    S=\frac{1}{16\pi G}\int\sqrt{-g}(R+L_{\mathrm{A}}+L_{\mathrm{M}})d^4x,
\end{equation}
where the aether Lagrangian is given by
\begin{equation}
    L_{A}=[-K^{ab}{}_{mn}\nabla_au^m\nabla_bu^n+\lambda(g_{ab}u^au^b+1)],
\end{equation}
and $K^{ab}{}_{mn}$ is defined as
\begin{equation}
    K^{ab}{}_{mn}=c_1g^{ab}g_{mn}+c_2\delta_m^a\delta_n^b+c_3\delta_n^a\delta_m^b-c_4u^au^bg_{mn}.
\end{equation}
Here, the $c_i$ are dimensionless coupling constants, and $\lambda$ is a Lagrange multiplier enforcing the unit timelike constraint on the aether.
The matter Lagrangian $L_{\mathrm{matter}}$ is assumed to depend on the metric tensor and the matter field only.

The field equations are obtained by extremizing the action.
The variation with respect to the Lagrange multiplier $\lambda$ imposes the condition that $u^a$ is a unit timelike vector
\begin{equation}
    u^a u_a = -1. \label{eq:ua}
\end{equation}
The variation of the action with respect to $u^a$ leads to the equation of motion of the aether field
\begin{equation}
    \nabla_aJ^a{}_b+c_4a_a\nabla_bu^a+\lambda u_b=0,\label{eq:EOMA}
\end{equation}
where 
\begin{equation}
    J^a{}_m=K^{ab}{}_{mn}\nabla_bu^n,
\end{equation}
and
\begin{equation}
    a_a=u^b\nabla_bu_a.
\end{equation}
The variation of the action with respect to the metric leads to the equation of motion of the gravitational field
\begin{equation}
    G_{ab}=8\pi G T_{ab}=T_{ab}^{A}+8\pi GT_{ab}^{M}, \label{eq:EOMg}
\end{equation}
where
\begin{equation}
\begin{aligned}
G_{ab}=&R_{ab}-\frac{1}{2}g_{ab}R,
\\
T_{ab}^{A}=&\nabla_{c}[J^{c}{}_{(a}u_{b)}+u^{c}J_{(ab)}-J_{(a}{}^{c}u_{b)}]\\
&-\frac{1}{2}g_{ab}J^{c}{}_{d}\nabla_{c}u^{d}+\lambda u_{a}u_{b}
\\
&+c_1[\nabla_au_c\nabla_bu^c-\nabla^cu_a\nabla_cu_b]+c_4a_aa_b,
\\
T_{ab}^{M}=&\frac{-2}{\sqrt{-g}}\frac{\delta\left(\sqrt{-g}L_{M}\right)}{\delta g_{ab}}.
\end{aligned}
\end{equation}
Contract Eq.~\eqref{eq:EOMA} with $u^b$ and use Eq.~\eqref{eq:ua}, we can solve the Lagrange multiplier $\lambda$ as
\begin{equation}
    \lambda = u^b\nabla_aJ^a{}_b+c_4a_a u^b\nabla_bu^a, \label{eq:sollambda}
\end{equation}
and substitute the solution into Eqs.~\eqref{eq:EOMA} and~\eqref{eq:EOMg}, we obtain the equations in terms of $u^a$ and $g_{ab}$ only.
For computational simplicity, we employ the trace-reversed equations derived from Eq.~\eqref{eq:EOMg}, which read
\begin{equation}
R_{ab} - \left(T_{ab}^A - \frac{1}{2} g_{ab} T^A\right) = 0. \label{eq:EOMg2}
\end{equation}

In the following, we shall solve the spherically symmetric static vacuum solutions with static aether configurations. 
That is, the only non-vanishing component of $u^a$ is $u^t$.
Under the static aether configuration, the vacuum solution depends exclusively on the parameter $c_{14}$. Consequently, we here provide a brief discussion on the broader physical significance of this coupling combination.
In the weak-field, slow-motion limit, the post-Newtonian analysis of the EA theory suggests an effective Newton’s constant as~\cite{Foster:2005dk}
\begin{equation}
    G=G_N\left(1-\frac{c_{14}}{2}\right),
\end{equation}
where $c_{14}=c_1+c_4$.
If $c_{14}=0$, the EA coupling constant G becomes the Newtonian coupling $G_N$.
For $c_{14}>2$, the coupling constant G becomes negative, implying that the gravity is repulsive. The coupling constant vanishes when $c_{14}=2$. 
Thus, a physically interesting region is $c_{14}< 2$ where the Newtonian limit can be recovered.
Furthermore, regarding gravitational wave polarizations in EA theory, the general parametric regime exhibits five distinct modes: two tensor, two vector, and one scalar polarization (2T+2V+1S)~\cite{Jacobson:2004ts}. Interestingly, in the specific limit where $c_{14} \to 0$, the vector and scalar sectors become degenerate or decouple, effectively reducing the active physical degrees of freedom to the two standard tensor polarizations, thereby recovering a GR-like propagation behavior.
For this parameter $c_{14}$, following the multi-messenger observations of GW170817~\cite{LIGOScientific:2017vwq} and its electromagnetic counterpart~\cite{LIGOScientific:2017zic}, supplemented by essential theoretical self-consistency conditions~\cite{Jacobson:2007veq}, its constraints have become exceptionally stringent. Specifically, these bounds are given by~\cite{Oost:2018tcv}
\begin{equation}
    0<c_{14}\lesssim2.5\times10^{-5}.
\end{equation}
Although experimental observations have severely constrained the parameter $c_{14}$, as a theoretical study, we analyze the vacuum solutions for all possible parametric regimes.

\section{Spherical Solutions of EA Field Equations with a Static Aether}\label{sec:sol}

\subsection{Preliminary Resolution of the Field Equations}

For a spherically symmetric vacuum solution, we adopt a general metric ansatz of the form
\begin{equation}
ds^2 = -G(\rho) dt^2 + H(\rho) d\rho^2 + R(\rho)^2 d\Omega^2,\label{eq:ds1}
\end{equation}
where $G(\rho)$, $H(\rho)$ and $R(\rho)$ are functions to be determined.
The field configuration for a static aether is $u^a = (u^t, 0,0,0)$.
In this field configuration, the condition that $u^a$ is a unit timelike vector requires that
\begin{equation}
    u^t = \frac{1}{\sqrt{G(\rho)}}.\label{eq:ut1}
\end{equation}
With the field configurations Eqs.~\eqref{eq:ds1} and~\eqref{eq:ut1} and the solution of the Lagrange multiplier Eq.~\eqref{eq:sollambda}, we can check that the equation of motion in the aether sector Eq.~\eqref{eq:EOMA} is identically satisfied.
Therefore, we only need to solve the equations in the gravity sector.
Denoting the left-hand side of Eq.~\eqref{eq:EOMg2} by $\mathcal{E}$, its individual components are explicitly given by 
\begin{equation}
\begin{aligned}
\mathcal{E}_{tt} = &
\frac{c_{14}-2}{8 G H^2 R}\Big(
G R G' H'+H R G'^2
\\&
-4 G H G' R'-2 G H R G''
\Big),
\\
\mathcal{E}_{\rho\rho} =&
\frac{1}{8 G^2 H R}\Big(
\left(3 c_{14}+2\right) H R G'^2
+8 G^2 H' R'
\\&+G' \left(\left(c_{14}+2\right) G R H'-4 c_{14} G H R'\right)
\\& -2 \left(c_{14}+2\right) G H R G'' 
-16 G^2 H R''
\Big)
\\
\mathcal{E}_{\theta\theta} = &
\frac{1}{8 G^2 H^2}\Big(
8 G^2 H^2 
+c_{14} H R^2 G'^2
+4 G^2 R H' R
\\&
+G' \left(c_{14} G R^2 H'-4 \left(c_{14}+1\right) G H R R'\right)
\\&
-2 c_{14} G H R^2 G''
-8 G^2 H R R''
-8 G^2 H R'^2
\Big)
\\
\mathcal{E}_{\phi\phi} =& \sin^2(\theta) \mathcal{E}_{\theta\theta},\label{eq:EOMs}
\end{aligned}
\end{equation}
where a prime indicates differentiation with respect to $\rho$.

As can be seen, for $c_{14}=2$, the component $\mathcal{E}_{tt}=0$ is automatically satisfied. Since this specific case has already been thoroughly investigated in Ref.~\cite{Chan:2020amr}, we shall not pursue it further here; henceforth, we assume $c_{14} \neq 2$.

Benefiting from diffeomorphism invariance, one can solve Eq.~\eqref{eq:EOMs} for other unknown functions by imposing a gauge choice on any single one of them. Intriguingly, under the ansatz $G(\rho) = \rho^\alpha$ (with $\alpha$ being a constant), the field equations are drastically simplified, allowing for exact analytical solutions for any $c_{14}$. Here, $\rho$ is understood as a dimensionless quantity, stripped of its spatial distance interpretation. Explicitly, within this configuration, $\mathcal{E}_{tt}=0$ reduces to
\begin{equation}
    H \left(4 \rho  R'+(\alpha -2) R\right)-\rho  R H'=0,
\end{equation}
and the solution for $H(\rho)$ is
\begin{equation}
    H(\rho) = K_1 \rho^{\alpha-2} R(\rho)^4,
\end{equation}
where $K_1$ is a constant.
Consequently, by fixing $\alpha=2$, the resulting solutions are further simplified, yielding
\begin{equation}
\begin{aligned}
    G(\rho) &= \rho^2,\\
    H(\rho) &= K_1 R(\rho)^4.\label{eq:GHrho}
\end{aligned}
\end{equation}
With Eq.~\eqref{eq:GHrho}, the equation $\mathcal{E}_{\rho\rho}=0$ reduces to
\begin{equation}
\frac{c_{14}}{\rho ^2}-\frac{2 R''}{R}+\frac{4 R'^2}{R^2}+\frac{2 R'}{\rho  R}=0,
\end{equation}
and the solution is 
\begin{equation}
R(\rho) = \frac{K_2 \rho^{\mu-1}}{\rho^{2\mu}-K_3},
\end{equation}
where $K_2$ and $K_3$ are constants, and
\begin{equation}
    \mu^2 = 1-c_{14}/2.
\end{equation}
As can be observed, the requirement that $\mu$ be a real number implies $c_{14} \le 2$, which precisely corresponds to the positivity condition of the Newton constant ($G_N \ge 0$). Furthermore, one can find that for the specific choices of $c_{14} = 3/2, 16/9, 48/25, -16, 2, 0$ in Ref.~\cite{Chan:2020amr}, $\mu$ reduces to a simple rational number in each case.

Up to this point, the field equation $\mathcal{E}_{\theta\theta} = 0$ remains to be verified. Under the aforementioned solutions, the component $\mathcal{E}_{\theta\theta} = 0$ reduces to
\begin{equation}
    \mathcal{E}_{\theta\theta} = 1-\frac{4\mu^2K_3}{K_1 K_2^2}=0.
\end{equation}
Therefore, $\mathcal{E}_{\theta\theta} = 0$ requires $K_1 = 4\mu^2 K_3/K_2^2$. 
Consequently, the full metric solution is explicitly given by
\begin{equation}
\begin{aligned}
G(\rho) &= \rho^2,\\
H(\rho) &= \frac{4 K_2^2 K_3 \mu ^2 \rho ^{4 (\mu -1)}}{\left(\rho ^{2 \mu }-K_3\right){}^4},\\
R(\rho) &= \frac{K_2 \rho ^{\mu -1}}{\rho ^{2 \mu }-K_3},\label{eq:pre_sol}
\end{aligned}
\end{equation}
where $K_2$ and $K_3$ are integration constants.

Although the metric seemingly depends on two parameters, $K_2$ and $K_3$, diffeomorphism invariance implies that one of them could be a consequence of gauge redundancy. Therefore, we proceed to investigate whether this configuration truly constitutes a physical two-parameter vacuum solution.

\subsection{Parameter Fixing through Asymptotic Behavior}

Although Eq.~\eqref{eq:pre_sol} presents concise and exact solutions, expressing them in terms of $\rho$ obscures the underlying physical picture, leaving the physical interpretations of the parameters $K_2$ and $K_3$ ambiguous. To address this, we perform a coordinate transformation to recast the metric into the standard form
\begin{equation}
 ds^2 = -g_{tt}(r) dt^2 + g_{rr}(r) dr^2 + r^2 d\Omega^2.
\end{equation}
Albeit an exact transformation is analytically intractable, we can alternatively carry out an asymptotic analysis in the far-field limit as $r \to \infty$.

The relations between $g_{tt}$, $g_{rr}$, $r$ and $\rho$ are given by
\begin{equation}
\begin{aligned}
r &= R(\rho) = \frac{K_2 \rho ^{\mu -1}}{\rho ^{2 \mu }-K_3},\\
g_{tt} &= G(\rho) = \rho^2, \\
g_{rr} &= H(\rho)\left(\frac{d\rho}{dr}\right)^2 = 
\frac{4 K_3 \mu ^2 \rho ^{2 \mu }}{\left((\mu -1)K_3 +(\mu +1) \rho ^{2 \mu }\right)^2}.
\end{aligned}
\end{equation}
We consider the region where $g_{tt}$ and $g_{rr}$ approach constants as $r \to \infty$. Evidently, this asymptotic regime implies $\rho^{2\mu} \to K_3$. Consequently, the asymptotic relation of $\rho$ with respect to $r$ can be solved as
\begin{equation}
\rho = K_3^{\frac{1}{2\mu}} +\frac{K_2}{2 \sqrt{K_3} \mu  r} -\frac{K_2^2 K_3^{-\frac{1}{2 \mu }-1}}{8 \mu ^2 r^2}
+\mathcal{O}\left(\frac{1}{r^3}\right).
\end{equation}
Hence, the asymptotic series of $g_{tt}$ and $g_{rr}$ are
\begin{equation}
\begin{aligned}
K_3^{-\frac{1}{\mu}}g_{tt}
= & 1+ \frac{K_2 K_3^{-\frac{\mu +1}{2 \mu }}}{\mu  r}-\frac{K_2^3 \left(\mu ^2-1\right) K_3^{-\frac{3 (\mu +1)}{2 \mu }}}{24 \mu ^3 r^3} 
\\& +\mathcal{O}\left(\frac{1}{r^4}\right),
\\
g_{rr} = & 1 -\frac{K_2 K_3^{-\frac{\mu +1}{2 \mu }}}{\mu  r}-\frac{K_2^2 \left(\mu ^2-5\right) K_3^{-\frac{\mu +1}{\mu }}}{4 \mu ^2 r^2}
\\& +\frac{K_2^3 \left(5 \mu ^2-13\right) K_3^{-\frac{3 (\mu +1)}{2 \mu }}}{8 \mu ^3 r^3} +\mathcal{O}\left(\frac{1}{r^4}\right).
\end{aligned}
\end{equation}
Therefore, we can identify the ADM mass to be
\begin{equation}
    M = -\frac{1}{2\mu} K_2 K_3^{-\frac{\mu +1}{2 \mu }}.
\end{equation}
Defining the constant $R_s = 2M$ and rescaling the time coordinate $t$ such that $g_{tt} \to 1$ as $r \to \infty$, we find that the asymptotic expansion of the metric simplifies to
\begin{equation}
\begin{aligned}
g_{tt} =& 1-\frac{R_s}{r}+\frac{\left(\mu ^2-1\right) R_s^3}{24 r^3}+\mathcal{O}\left(\frac{1}{r^4}\right),\\
g_{rr} =& 1+\frac{R_s}{r}-\frac{\left(\mu ^2-5\right) R_s^2}{4 r^2}-\frac{\left(5 \mu ^2-13\right) R_s^3}{8 r^3}\\&+\mathcal{O}\left(\frac{1}{r^4}\right).\label{eq:infasymptotic}
\end{aligned}
\end{equation}
It is worth noting that $\mu$ appears exclusively in even powers within this asymptotic expansion, meaning that both the positive and negative signs of $\mu$ correspond to the exact same solution. Consequently, we can restrict our consideration to the positive branch with $\mu \geq 0$, which yields
\begin{equation}
    \mu = \sqrt{1-c_{14}/2}.
\end{equation}

Evidently, this asymptotic expansion demonstrates that the metric is uniquely determined by the ADM mass $M = R_s/2$; thus, only one of the two parameters, $K_2$ and $K_3$, is physically independent. Here, we choose to fix $K_3 = 1$ and express $K_2$ in terms of $R_s$. Under this gauge fixing, the complete metric is expressed as
\begin{equation}
\begin{aligned}
ds^2 =& -\rho^2 dt^2 + \frac{4 \mu ^4 R_s^2}{\rho ^4 \left(\rho ^{-\mu }-\rho ^{\mu }\right)^4} d\rho^2\\&
+\frac{\mu ^2 R_s^2}{\rho ^2 \left(\rho ^{-\mu }-\rho ^{\mu }\right)^2}(d\theta^2+\sin^2\theta d\phi^2). \label{eq:gsol}
\end{aligned}
\end{equation}
The relations between $g_{tt}$, $g_{rr}$, $r$ and $\rho$ now reduce to
\begin{equation}
\begin{aligned}
g_{tt} &= \rho^2,\\
g_{rr} &= \frac{4 \mu ^2}{\left((\mu -1) \rho ^{-\mu }+(\mu +1) \rho ^{\mu }\right)^2},\\
r &= \frac{\mu  R_s}{\rho  \left(\rho ^{-\mu }-\rho ^{\mu }\right)}. \label{eq:metinr}
\end{aligned}
\end{equation}
It is straightforward to verify that for $c_{14}=0$, we have $\mu = 1$ and $\rho^2 = 1 - R_s/r$. Consequently, the solution smoothly reduces to the standard Schwarzschild solution.

\begin{figure}[htb]
	\centering
	\includegraphics[width=0.75 \linewidth]{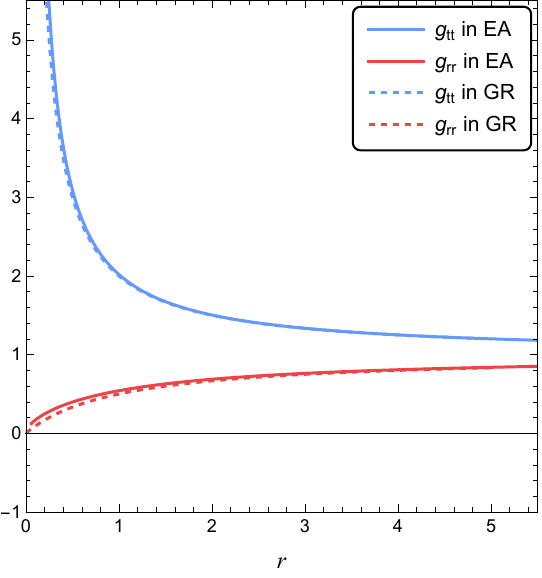}
	\caption{Plot of the metric components $g_{tt}$ and $g_{rr}$ against the radius $r$ for $\mu = 1/3$ ($c_{14} = 16/9$) and $R_s = -1$. Solid and dashed curves denote the solutions in EA theory and GR, respectively. The condition $R_s < 0$ denotes an unphysical parameter region.}
	\label{fig:unphysical}
\end{figure}

For the solution to be physically viable, the ADM mass must be positive ($R_s > 0$), which yields $0 < \rho < 1$ for a positive $r$. In the unphysical case where $R_s < 0$, we have $\rho > 1$. Furthermore, the condition $\rho > 1$ implies that both $g_{tt}$ and $g_{rr}$ are free of poles, which indicates that this unphysical solution possesses neither an event horizon nor a wormhole throat. The corresponding behaviors of $g_{tt}$ and $g_{rr}$ are depicted in Fig.~\ref{fig:unphysical}. In fact, solutions of this specific form were previously obtained in Ref.~\cite{Chan:2020amr} (explicitly, their $c_{14}=16/9$ case), displaying profiles identical to those in Fig.~\ref{fig:unphysical}.

To locate the singularities of the obtained solution, we calculate the Kretschmann scalar $K\equiv R_{abcd}R^{abcd}$. The explicit computation shows that
\begin{equation}
    K = \frac{\rho ^{4-8 \mu } \left(\rho ^{2 \mu }-1\right)^6}{4 \mu ^8 R_s^4}(A+B \rho^{2\mu} +C \rho^{4\mu}), \label{eq:Kre}
\end{equation}
where
\begin{equation}
\begin{aligned}
    A = & (\mu -1)^2 \left(3 \mu ^2-2 \mu +7\right),\\
    B = & -6 \mu ^4+20 \mu ^2-14,\\
    C = & (\mu +1)^2 \left(3 \mu ^2+2 \mu +7\right)
\end{aligned}
\end{equation}
are constants.
Given that $\mu \ge 0$, the boundary $\rho \to \infty$ emerges as a potential singularity where $g_{tt} \to \infty$. Furthermore, for $\mu > 1/2$, the point $\rho = 0$ also constitutes a singularity, at which $g_{tt} = 0$, corresponding to the parameter space $c_{14} < 3/2$.

\section{Geometries across Parameter Spaces}\label{sec:geo}

In this section, we analyze the geometric topology of the spacetime described by the metric \eqref{eq:gsol} for various choices of $c_{14}$.
This analysis stems from Eq.~\eqref{eq:metinr}, from which we can clearly observe that for $\mu < 1$ ($c_{14} > 0$), the metric component $g_{rr}$ exhibits a pole. The location of this pole is given by
\begin{equation}
\begin{aligned}
    \rho_0 = & \left(\frac{1-\mu}{1+\mu}\right)^\frac{1}{2\mu},\\
    r_0 = & \frac{1}{2} (1+\mu)^\frac{1+\mu}{2\mu} (1-\mu)^\frac{-1+\mu}{2\mu}R_s.
\end{aligned}
\end{equation}
On the other hand, the results for the curvature scalars imply that this pole is merely a coordinate singularity, and the spacetime remains well-defined at this location.
However, $g_{tt} \neq 0$ at this point, implying that the configuration does not represent a black hole solution but rather points to a wormhole-like structure.
For $\mu > 1$ ($c_{14} < 0$), no pole appears in $g_{rr}$ for $\rho \neq 0$. Given that the singularity, as analyzed above, can only potentially occur at $\rho = 0$, this spacetime likely describes a naked singularity. Below, we proceed to investigate these two scenarios in detail.

\subsection{\texorpdfstring{The Case of $\mu > 1$ ($c_{14} < 0$)}{The Case of mu > 1 (c14 < 0)}}

\begin{figure}[htb]
	\centering
	\includegraphics[width=0.75\linewidth]{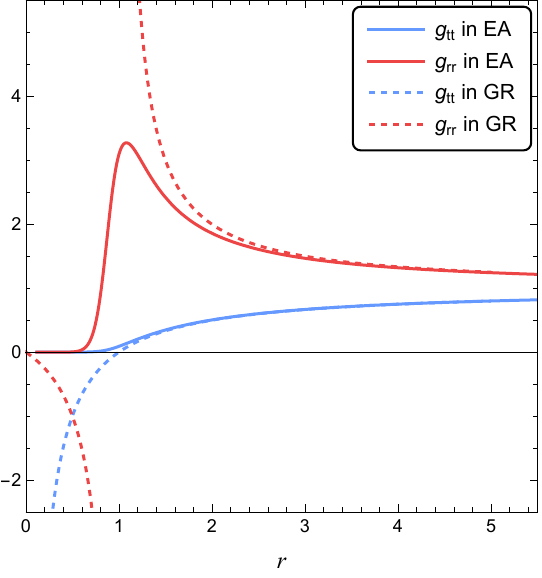}
	\caption{Plot of the metric components $g_{tt}$ and $g_{rr}$ against the radius $r$ for $\mu = 6/5$ ($c_{14} = -22/25$) and $R_s = 1$. Solid and dashed curves denote the solutions in EA theory and GR, respectively.}
	\label{fig:2}
\end{figure}

We first consider the case $\mu > 1$. In this parameter range, $g_{tt}$ and $g_{rr}$ possess no singularities and exhibit zeros solely at $\rho = 0$, and thus the geometry is devoid of horizons and wormhole throats. Fig.~\ref{fig:2} depicts the profiles of $g_{tt}$ and $g_{rr}$ with respect to $r$.
As illustrated in the figure, $\rho = 0$ corresponds to $r = 0$, where both $g_{tt}$ and $g_{rr}$ vanish. Consequently, the point $r = 0$ constitutes a naked singularity.

The asymptotic behavior of $\rho(r)$ near $\rho = 0$ can also be obtained via a series expansion, expressed as
\begin{equation}
\rho(r) = \left( \frac{r}{R_0} \right)^{\frac{1}{\mu-1}} - \frac{1}{\mu-1} \left( \frac{r}{R_0} \right)^{\frac{2\mu+1}{\mu-1}} + \mathcal{O}\left( r^{\frac{4\mu+1}{\mu-1}} \right),
\end{equation}
where $R_0 = \mu R_s$.
Consequently, the asymptotic expansion of the metric in the vicinity of $r \to 0$ can be approximated as
\begin{equation}
    ds^2\simeq -\left(\frac{R_0}{r}\right)^{b-2} dt^2 + b^2 \left(\frac{R_0}{r}\right)^{b} dr^2 +r^2 d\Omega^2,\label{eq:met0}
\end{equation}
where $b=2\mu/(\mu-1)>2$.
Remarkably, the asymptotic metric in Eq.~\eqref{eq:met0} is formally identical in structure to Eq.~\eqref{eq:metinf}, which was previously obtained for the $r \to \infty$ limit in the $\mu < 1$ regime.

\subsection{\texorpdfstring{The Case of $\mu < 1$ ($c_{14} > 0$)}{The Case of mu < 1 (c14 > 0)}}

\begin{figure}[htb]
	\centering
	\includegraphics[width=0.75\linewidth]{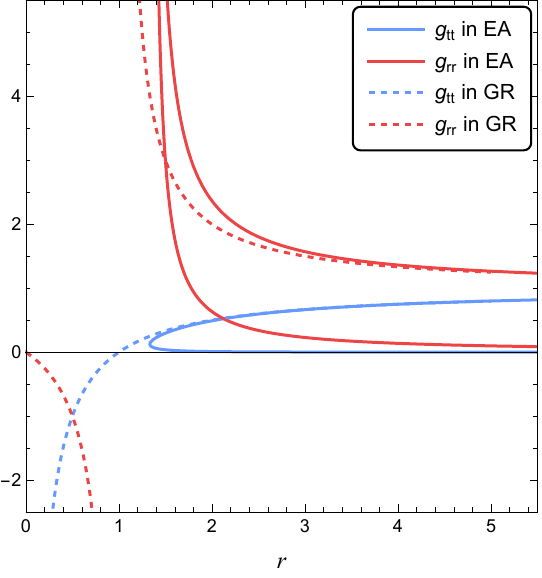}
	\caption{Plot of the metric components $g_{tt}$ and $g_{rr}$ against the radius $r$ for $\mu = 1/3$ ($c_{14} = 16/9$) and $R_s = 1$. Solid and dashed curves denote the solutions in EA theory and GR, respectively.}
	\label{fig:1}
\end{figure}

In the case where $\mu < 1$, Eq.~\eqref{eq:metinr} allows us to illustrate the behaviors of $g_{rr}$ and $g_{tt}$ as functions of $r$, as shown in Fig.~\ref{fig:1}.
Within this parameter regime, it is apparent that both $g_{tt}$ and $g_{rr}$ remain positive everywhere, with no sign reversal occurring.
Additionally, the metric exhibits completely different behaviors on either side of the pole $r = r_0$ of $g_{rr}$. Specifically, the metric is undefined for $r < r_0$, while for any $r > r_0$, the metric components $g_{tt}$ and $g_{rr}$ are double-valued.
Such a structure indicates that $r = r_0$ resembles a wormhole throat, through which a test particle reaching this radius can cross into another spacetime domain.

However, the geometries of these two spacetime regions are starkly different. As can be inferred from Fig.~\ref{fig:1}, in one region, the metric components satisfy $g_{tt}, g_{rr} \to 1$ as $r \to \infty$, indicating that this domain is asymptotically flat, with its asymptotic expansion given by Eq.~\eqref{eq:infasymptotic}. Conversely, in the other spacetime region, we observe that $g_{tt}$ and $g_{rr}$ approach zero as $r \to \infty$; hence, this region does not constitute an asymptotically flat spacetime.
Therefore, we proceed to investigate the asymptotic behavior of the other spacetime branch in the limit $r \to \infty$.

From the relationship between $r$ and $\rho$ defined in Eq.~\eqref{eq:metinr}, one can asymptotically solve for the alternative branch of $\rho(r)$ in the limit $r \to \infty$ within the region $\mu < 1$. The explicit result is given by
\begin{equation}
\rho(r) = \left(\frac{R_0}{r}\right){}^{\frac{1}{1-\mu }}+\frac{1}{1-\mu }\left(\frac{R_0}{r}\right){}^{\frac{2 \mu +1}{1-\mu }}+ \mathcal{O}\left( r^{\frac{4\mu+1}{\mu-1}} \right),
\end{equation}
where $R_0=\mu R_s$.
Concomitantly, we can establish the asymptotic expansions for the metric as $r \to \infty$, which can be approximated as
\begin{equation}
ds^2\simeq -\left(\frac{R_0}{r}\right)^{2+a} dt^2 + a^2 \left(\frac{R_0}{r}\right)^{a} dr^2 +r^2 d\Omega^2,\label{eq:metinf}
\end{equation}
where $a = 2\mu/(1-\mu) > 0$.
As $r \to \infty$, the Kretschmann scalar exhibits the following asymptotic behavior:
\begin{equation}
K=R_{abcd}R^{abcd}\simeq C r^{2\alpha-4},
\end{equation}
Thus, for $a > 2$, the infinity $r \to \infty$ corresponds to a physical singularity, and this consequently implies that
\begin{equation}
    \frac{1}{2}<\mu<1,\quad 0<c_{14}<\frac{3}{2}.
\end{equation}
This result can also be deduced directly from the original metric. In fact, we have already demonstrated that $\rho = 0$ corresponds to a curvature singularity when $\mu > 1/2$, which precisely maps to the boundary at infinity for this specific spacetime region.
It seems that the boundary is not a curvature singularity when $\mu \leq 1/2$.
However, as argued in Ref.~\cite{Eling:2006df}, this boundary remains a physical singularity due to the divergence of $R_{ab} k^a k^b$, with $k^a$ being the tangent vector of an affinely parameterized radial null geodesic approaching the internal infinity.

We now examine the finiteness of the radial length and volume for this spacetime domain. The proper distance extending from $r = r_1$ to $r \to \infty$ is given by
\begin{equation}
    L \simeq a \int_{r_1}^\infty \left(\frac{R_0}{r}\right)^{a/2} dr.
\end{equation}
Therefore, requiring the proper distance to be finite yields the constraint $a > 2$.
This parameter range exactly matches the condition for the singularity at $r \to \infty$. Therefore, the geometric structure of this region implies that despite being situated at $r \to \infty$, the singularity is at a finite proper distance from the throat. Moreover, this singularity is a space-like surface possessing an infinite area.
Conversely, when $0 < \mu \le 1/2$ (corresponding to $3/2 \le c_{14} < 2$), the boundary at $r \to \infty$ no longer constitutes a singularity, and its proper distance from the wormhole throat becomes infinite.

Regarding the volume extending from $r = r_1$ to $r \to \infty$, its explicit expression is formulated as
\begin{equation}
    V\simeq 4\pi a \int_{r_1}^\infty \left(\frac{R_0}{r}\right)^{a/2} r^2 dr.
\end{equation}
Consequently, for $a > 6$, the convergence of the integral indicates that even the volume of this spacetime domain is finite, which restricts the parameters to
\begin{equation}
    \frac{3}{4}<\mu<1,\quad 0<c_{14}<\frac{7}{8}.
\end{equation}

\section{Photon Spheres and Geodesic Completeness}\label{sec:PSGC}

In this section, we utilize the analytical metric expressed in Eq.~\eqref{eq:gsol} to perform a detailed investigation of the geodesic behaviors. In particular, we focus our attention on the critical properties of photon spheres and the assessment of geodesic completeness across different parameter spaces.

\subsection{Photon Spheres and Naked Singularities}

A photon sphere is defined to be a nowhere-spacelike hypersurface, such that any photon whose 4-velocity is initially tangent to the surface will always remain in the surface if not altered~\cite{Claudel:2000yi}.
More intuitively, as a light ray approaches the photon sphere, the Einstein bending angle tends to be infinitely large~\cite{Virbhadra:1999nm, Virbhadra:2002ju}.
For a Schwarzschild black hole, the gravitational lensing would cause relativistic images due to photon spheres, besides the ordinary primary and secondary images~\cite{Virbhadra:1999nm}.
Furthermore, the existence or non-existence of photon spheres around a singularity~(naked or not) has important implications for gravitational lensing, which can be compared to realistic observations.
In the literature, a singularity with or without photon spheres is termed a  {\it weakly} or {\it strongly naked singularity}~(WNS or SNS) respectively.

For the spacetime metric given in Eq.~\eqref{eq:ds1}, the radius of the photon sphere is determined by
\begin{equation}
    \frac{d}{d\rho} \left(\frac{G(\rho)}{R(\rho)^2}\right)=0.
\end{equation}
Aside from the trivial solutions $\rho = 0$ and $\rho = 1$ (which correspond to $r = 0$ or $r \to \infty$), for $\mu \geq 2$, the photon sphere ceases to exist. When $\mu<2$, the above equation yields the position of the photon sphere at
\begin{equation}
    \rho_{\text{ph}} = \left(\frac{2-\mu}{2+\mu}\right)^\frac{1}{2\mu},
\end{equation}
Accordingly, the corresponding radius $r^*$ is
\begin{equation}
    r_{\text{ph}} = \frac{1}{2} (2+\mu)^\frac{1+\mu}{2\mu}(2-\mu)^\frac{\mu-1}{2\mu}R_s.
\end{equation}
It is straightforward to verify that in the limit $\mu \to 1$, the photon sphere radius approaches $r_{\text{ph}} \to 3 R_s/2$, which precisely recovers the classical GR result where the photon sphere is located at $3M$.

\begin{figure}[htb]
	\centering
	\includegraphics[width=0.8\linewidth]{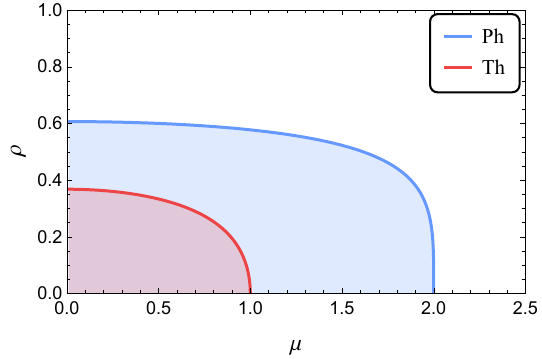}
	\caption{Plot of the photon sphere position (Ph) and the throat position (Th) against the parameter $\mu$ for $R_s = 1$. The abscissa and ordinate denote the parameter $\mu$ and the radial distance, respectively.}
	\label{fig:phth}
\end{figure}

Fig.~\ref{fig:phth} illustrates the behavior of the photon sphere position $\rho_{\text{ph}}$ and the throat location $\rho_{\text{th}}$ as functions of the parameter $\mu$, where a larger value of $\rho$ corresponds to a region closer to the spatial infinity ($r \to \infty$) of the asymptotically flat spacetime. It is manifest that within the range $0 < \mu < 1$, the photon sphere completely envelops the wormhole structure.
In the intermediate regime $1 < \mu < 2$, the photon sphere instead cloaks the singularity, rendering it a WNS. Intriguingly, when $\mu \ge 2$, the photon sphere ceases to exist, leaving the curvature singularity entirely exposed as an SNS.

\subsection{Geodesic Completeness}

Here, we examine the geodesic completeness.
For a test particle moving in the spacetime metric \eqref{eq:ds1}, assuming the motion takes place in the $\theta = \pi/2$ plane, the underlying symmetries allow us to construct the corresponding constants of motion
\begin{equation}
\begin{aligned}
    E = & G(\rho) \dot{t},\\
    L = & R(\rho)^2 \dot{\phi},
\end{aligned}
\end{equation}
where $E$ represents the energy and $L$ signifies the angular momentum.
In addition, the four-velocity of the particle satisfies the normalization condition
\begin{equation}
-G \dot{t}^2 +H \dot{\rho}^2+R^2 \dot{\phi}^2=-\epsilon,
\end{equation}
where $\epsilon = 1$ and $\epsilon = 0$ correspond to massive and massless particles, respectively.
Consequently, we can establish the radial equation of motion for the particle as
\begin{equation}
    \dot{\rho}^2 = \frac{1}{H}\left(\frac{E^2}{G}-\frac{L^2}{R^2}-\epsilon\right).
\end{equation}
Considering a particle moving along the axial direction with $L = 0$, its substitution into the metric \eqref{eq:gsol} leads to
\begin{equation}
    \frac{d\rho}{d\tau} = \pm \frac{\rho^{1-2\mu}}{2\mu^2 R_s}(\rho^{2\mu}-1)^2\sqrt{E^2-\epsilon \rho^2},
\end{equation}
where $\tau$ denotes either the proper time of the particle or an affine parameter along the geodesic.
Consider a particle moving from $\rho = \rho_1 \ll 1$ in the direction of $\rho \to 0$, which implies $\rho^{2\mu} \ll 1$ and $\rho \ll 1$. For both massive and massless particles, we consistently obtain
\begin{equation}
    \tau_{tot} \simeq \frac{2\mu R_s^2}{E}\int_{0}^{\rho_1} \rho^{2\mu-1} d\rho = \frac{\mu R_s}{E}\rho_1^{2\mu}.
\end{equation}
Consequently, for any arbitrary parameter $\mu$, regardless of whether the particle is massive or massless, it will inevitably reach the singularity at $\rho = 0$ within a finite proper time or affine parameter.
For $\mu > 1/2$, the boundary $\rho = 0$ constitutes a curvature singularity, implying that the spacetime in this regime is geodesically incomplete.
In the range $0 < \mu \le 1/2$, where $\rho = 0$ corresponds to the spatial infinity of the other spacetime region, particles are found to reach infinity in a finite proper time or affine parameter, which highlights the geodesic incompleteness of the spacetime.

Crucially, from the perspective of a distant observer at spatial infinity within the asymptotically flat spacetime branch, the coordinate time required for this particle's motion is
\begin{equation}
    T \simeq 2\mu^2 R_s \int_{0}^{\rho_1}\rho^{2\mu-3}d\rho.
\end{equation}
The convergence of this integral strictly requires $\mu > 1$. Structurally, when the spacetime manifests as a naked singularity with $\mu > 1$, an observer at asymptotic infinity will witness the particle striking the singularity within a finite coordinate time. Conversely, within the regime $0 < \mu < 1$ where the spacetime exhibits a multi-sheeted geometry connected via a wormhole throat, the distant observer will instead find that the particle asymptotically freezes, never managing to reach either the alternative singularity or the adjacent spatial infinity in any finite duration.

\section{Analytic Extension of the Solution}\label{sec:Analytic}

Although the analytical solution has been derived in the standard $(t, r, \theta, \phi)$ coordinates as shown in Eq.~\eqref{eq:metinr}, it restrictively describes the branch with $g_{tt} > 0$ when the parameter $\rho$ takes real values. Consequently, an analytic extension of this solution is indispensable to ascertain the potential existence of a branch characterized by $g_{tt} < 0$.

Due to the presence of the exponent $\mu$ in the metric components, a direct analytic continuation via the transformation $\rho \to i \rho$ entails subtle branching ambiguities, necessitating a rigorous treatment of the branch cuts to ensure the continuity of the metric across the complex plane.
Instead of analytically extending the solution \eqref{eq:gsol}, we present a more direct method by solving the equations from scratch. In particular, in analogy to Eq.~\eqref{eq:gsol}, we employ the configuration $G(\rho) = -\rho^2$ and $H(\rho)<0$ to solve the governing equations, which yields the following exact solution:
\begin{equation}
\begin{aligned}
    G(\rho) = & -\rho^2,\\
    H(\rho) = & -\frac{4 \mu ^4 R_s^2}{\rho ^4 \left(\rho ^{-\mu }+\rho ^{\mu }\right)^4},\\
    R(\rho) = & \frac{\mu R_s}{\rho \left(\rho ^{-\mu }+\rho ^{\mu }\right)},
\end{aligned}
\end{equation}
where $\mu=\sqrt{1-c_{14}/2}$.
Hence, the metric is
\begin{equation}
\begin{aligned}
ds^2 =& \rho^2 dt^2 - \frac{4 \mu ^4 R_s^2}{\rho ^4 \left(\rho ^{-\mu }+\rho ^{\mu }\right)^4} d\rho^2\\&
+\frac{\mu ^2 R_s^2}{\rho ^2 \left(\rho ^{-\mu }+\rho ^{\mu }\right)^2} 
(d\theta^2+\sin^2\theta d\phi^2). \label{eq:gsol2}
\end{aligned}
\end{equation}
Correspondingly, under the standard spherically symmetric coordinates with the line element $ds^2 = -g_{tt}(r) dt^2 + g_{rr}(r) dr^2 + r^2 d\Omega^2$, we have
\begin{equation}
\begin{aligned}
g_{tt} &= -\rho^2,\\
g_{rr} &= -\frac{4 \mu ^2}{\left((1-\mu) \rho ^{-\mu }+(\mu +1) \rho ^{\mu }\right)^2},\\
r &= \frac{\mu  R_s}{\rho  \left(\rho ^{-\mu }+\rho ^{\mu }\right)}. \label{eq:metinr2}
\end{aligned}
\end{equation}

This geometry is entirely distinct from that of Eq.~\eqref{eq:gsol}. Specifically, one can observe that for $\mu > 1$, the metric component $g_{rr}$ of the present solution exhibits a pole.
One can easily verify that when $\mu = 1$, we have $g_{tt} = g_{rr} = 1 - R_s/r$, and the metric returns to the Schwarzschild solution. Remarkably, since this solution clearly differs from Eq.~\eqref{eq:gsol}, it constitutes another one-parameter family of solutions that overlaps with Eq.~\eqref{eq:gsol} only at $\mu = 1$ to retrieve the Schwarzschild geometry.
The Kretschmann invariant for the present solution is analogous to Eq.~\eqref{eq:Kre} under the substitution $\rho \to i\rho$. This implies that the true curvature singularities still emerge as $\rho \to \infty$, or at $\rho = 0$ when $\mu > 1/2$.

\begin{figure}[htb]
	\centering
	\includegraphics[width=0.75\linewidth]{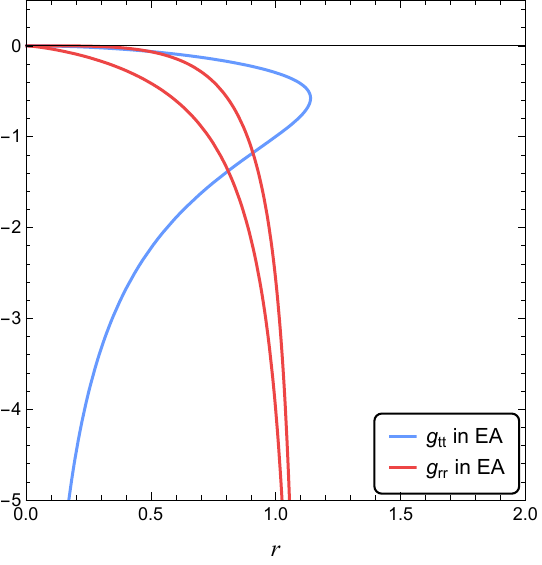}
	\caption{Plot of the metric components $g_{tt}$ and $g_{rr}$ against the radius $r$ for $\mu = 2$ ($c_{14} = -6$) and $R_s = 1$, where the metric components are from Eq.~\eqref{eq:metinr2}}
	\label{fig:muge1}
\end{figure}

We now discuss the parameter space by classifying the values of $\mu$. When $\mu > 1$, the singularity of $g_{rr}$ emerges at
\begin{equation}
\begin{aligned}
    \rho_0 = & \left(\frac{\mu-1}{\mu+1}\right)^\frac{1}{2\mu},\\
    r_0 = & \frac{1}{2} (\mu+1)^\frac{\mu+1}{2\mu} (\mu-1)^\frac{\mu-1}{2\mu}R_s.
\end{aligned}
\end{equation}
The profiles of $g_{tt}$ and $g_{rr}$ with respect to the radius $r$ for this scenario are illustrated in Fig.~\ref{fig:muge1}. As clearly manifested, the spacetime terminates at a maximum physical radius $r = r_0$.
This configuration seems to be an $S^4$-like topology: the spatial volume expands until it attains a maximum radius at $r = r_0$, subsequently undergoing a geometric contraction. Moreover, the two asymptotic regimes where $r = 0$ correspond to $\rho = 0$ and $\rho \to \infty$, respectively, indicating that both boundaries culminate in genuine curvature singularities.
However, given that $g_{rr} < 0$ in this metric, the radial coordinate $r$ effectively functions as a timelike coordinate. Consequently, unless a reversal of the arrow of time is permitted, the mapping of a single value of $r$ onto two distinct branches lacks a clear physical interpretation.
Given that $\rho = 0$ marks a curvature singularity instead of a null Killing horizon, this solution should be properly interpreted as a cosmological-like model rather than an analytical extension of a naked singularity. In this perspective, the physical interpretation becomes highly clear: $\rho$ plays the role of a cosmological time coordinate, where the curvature singularities at $\rho = 0$ and $\rho \to \infty$ imply that the universe undergoes a finite evolutionary lifespan, beginning from an initial singularity and ending at a final singularity.

\begin{figure}[htb]
	\centering
	\includegraphics[width=0.75\linewidth]{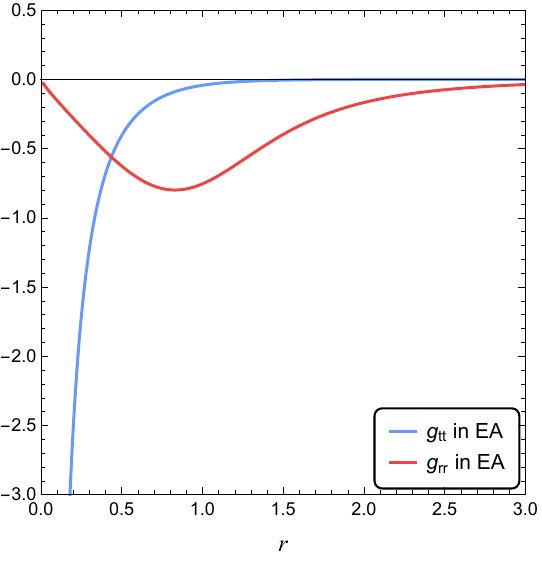}
	\caption{Plot of the metric components $g_{tt}$ and $g_{rr}$ against the radius $r$ for $\mu = 2/3$ ($c_{14} = 10/9$) and $R_s = 1$, where the metric components are from Eq.~\eqref{eq:metinr2}}
	\label{fig:mule1}
\end{figure}

We next consider the case of $\mu < 1$, for which the behavior of $g_{tt}$ and $g_{rr}$ as functions of $r$ is depicted in Fig.~\ref{fig:mule1}.
As can be seen, the radial coordinate in this scenario spans the entire range of $r \in (0, \infty)$. Specifically, as $r \to \infty$, we have $g_{tt} \to 0$ and $\rho \to 0$, whereas the limit $r \to 0$ leads to $g_{tt} \to -\infty$ and $\rho \to \infty$. Consequently, for the parameter range $1/2 < \mu < 1$, both the asymptotic infinity $r \to \infty$ and the central core $r = 0$ constitute genuine curvature singularities. In contrast, when $\mu \le 1/2$, the curvature singularity emerges exclusively at the center $r = 0$.

We now consider the asymptotic expansions of the metric as $r \to \infty$ and $r \to 0$. For the asymptotic infinity $r \to \infty$, we have
\begin{equation}
    \rho(r) = \left(\frac{R_0}{r}\right){}^{\frac{1}{1-\mu }}-\frac{1}{1-\mu }\left(\frac{R_0}{r}\right){}^{\frac{2 \mu +1}{1-\mu }}+ \mathcal{O}\left( r^{\frac{4\mu+1}{\mu-1}} \right),
\end{equation}
where $R_0 = \mu R_s$. Correspondingly, the metric is
\begin{equation}
ds^2\simeq \left(\frac{R_0}{r}\right)^{2+a_1} dt^2 - a_1^2 \left(\frac{R_0}{r}\right)^{a_1} dr^2 +r^2 d\Omega^2,\label{eq:metinf2}
\end{equation}
where $a_1 = 2\mu/(1-\mu) > 0$.
It is worth noting that this leading-order metric differs from Eq.~\eqref{eq:metinf} only in that the signs in front of $dt^2$ and $dr^2$ are exactly opposite.
Conversely, near the central singularity as $r \to 0$, we have
\begin{equation}
\rho(r) = \left(\frac{R_0}{r}\right){}^{\frac{1}{1+\mu }}-\frac{1}{1+\mu }\left(\frac{R_0}{r}\right){}^{\frac{1-2\mu}{1+\mu }}+ \mathcal{O}\left( r^{\frac{1-4\mu}{\mu+1}} \right),
\end{equation}
and the corresponding metric is
\begin{equation}
    ds^2\simeq \left(\frac{R_0}{r}\right)^{2-a_2} dt^2 - a_2^2 \left(\frac{r}{R_0}\right)^{a_2} dr^2 +r^2 d\Omega^2,
\end{equation}
where $a_2 = 2\mu/(1+\mu)$ and $0<a_2<1$.
Evidently, by taking the limit $\mu \to 1$, this metric smoothly reduces to the asymptotic behavior of the standard Schwarzschild solution in the near-singularity limit where $r \to 0$.

In Ref.~\cite{Eling:2006df}, Jacobson et al. identified a ``would-be'' Killing horizon at asymptotic infinity for the solution branch penetrating the wormhole under $0 < c_{14} < 2$. 
Here, we explicitly demonstrate that this singularity is actually an extreme Killing horizon, where the regions on either side smoothly match the metric configurations given in Eq.~\eqref{eq:gsol} and Eq.~\eqref{eq:gsol2}.

We first verify that, for both metrics, the asymptotic infinity corresponding to $\rho = 0$ constitutes an extremal Killing horizon within the parameter range $0 < \mu < 1$.
Regarding the metric of Eq.~\eqref{eq:gsol}, with the Killing vector $\xi^\mu = (1,0,0,0)$ and the surface normal $n_\mu = (0,1,0,0)$, we have
\begin{equation}
\begin{aligned}
\xi^\mu \xi_\mu &=-\lim_{\rho\to 0} \rho^2 = 0,\\
n_\mu n^\mu &= \lim_{\rho\to 0}\frac{\rho^{4-4\mu}(\rho^{2\mu}-1)^4}{4\mu^4R_s^2}= 0,
\end{aligned}
\end{equation}
Therefore, $\rho=0$ is a null hypersurface on which the Killing vector field vanishes, confirming that it is indeed a Killing horizon.
The surface gravity can be obtained by
\begin{equation}
\begin{aligned}
    \kappa^2 =& -\frac{1}{2} (\nabla_\mu \xi_\nu)(\nabla^\mu \xi^\nu)\\
    =& \lim_{\rho\to 0}\frac{\rho^{4-4\mu}(\rho^{2\mu}-1)^4}{4\mu^4R_s^2}= 0.
\end{aligned}
\end{equation}
Therefore, the boundary $\rho = 0$ is indeed an extremal Killing horizon.
For the metric Eq.~\eqref{eq:gsol2}, we have
\begin{equation}
\begin{aligned}
\xi^\mu \xi_\mu &=\lim_{\rho\to 0} \rho^2 = 0,\\
n_\mu n^\mu &= -\lim_{\rho\to 0}\frac{\rho^{4-4\mu}(\rho^{2\mu}+1)^4}{4\mu^4R_s^2}= 0,
\end{aligned}
\end{equation}
and
\begin{equation}
\begin{aligned}
    \kappa^2 =& -\frac{1}{2} (\nabla_\mu \xi_\nu)(\nabla^\mu \xi^\nu)\\
    =& \lim_{\rho\to 0}\frac{\rho^{4-4\mu}(\rho^{2\mu}+1)^4}{4\mu^4R_s^2}= 0.
\end{aligned}
\end{equation}
So for the metric of Eq.~\eqref{eq:gsol2}, the boundary $\rho \to 0$ is also verified to be an extremal Killing horizon.
Notably, the components $g_{tt}$ and $g_{rr}$ undergo a signature inversion across the horizon at $\rho = 0$. This sign reversal is a characteristic signature of a coordinate-independent causal boundary, where the roles of the time and radial coordinates are effectively interchanged.

To further investigate whether a test particle originating from an asymptotically flat region can successfully traverse the horizon located at $\rho = 0$, we perform an analysis within the framework of Kruskal-like coordinates. Starting from the explicit metric form in Eq.~\eqref{eq:gsol}, we obtain
\begin{equation}
    ds^2 = -\rho^2 du dv +R^2 d\Omega^2,
\end{equation}
where
\begin{equation}
\begin{aligned}
u = & t-\rho*,\\
v = & t+\rho*,\\
\rho* 
=& -\frac{\mu^2 R_s}{1-\mu}\rho^{-2(1-\mu)} \, _2F_1\left(2,1-\frac{1}{\mu };2-\frac{1}{\mu };\rho ^{2 \mu }\right).
\end{aligned}
\end{equation}
Specifically, as $\rho \to 1$, the tortoise coordinate diverges to $+\infty$, whereas as $\rho \to 0$, it approaches $-\infty$. Consequently, for an infalling particle propagating toward the horizon at $\rho = 0$, the advanced null coordinate $v$ remains finite, while the retarded null coordinate $u$ diverges to $+\infty$.
Near $\rho = 0$, we have $\rho* \simeq -C_1 \rho^{-2(1-\mu)}$, where $C_1$ is a constant, and consequently, the metric can be approximated as
\begin{equation}
ds^2 \simeq - C_2 \frac{du dv}{(u-v)^\gamma} +R^2 d\Omega^2,
\end{equation}
where $C_2$ is a constant, and $\gamma = 1/(1-\mu)>1$.
Because the surface gravity vanishes near $\rho = 0$, the standard exponential Kruskal coordinates are no longer applicable. Instead, we introduce a power-law type Kruskal parameterization to analyze the horizon geometry~\cite{Liberati:2000sq}, which reads explicitly as
\begin{equation}
\begin{aligned}
    u = & U^{-\alpha},\\
    v = & -V^{-\alpha},
\end{aligned}
\end{equation}
where $\alpha = 1/\mu -1>0$.
Under this parameterization, the limit $u \to +\infty$ corresponds to $U = 0$, and a finite value of $v$ yields a finite $V$. The metric becomes
\begin{equation}
    ds^2 \simeq \alpha^2 C_2 \frac{dU dV}{\left(U^\alpha+V^\alpha\right)^\gamma} + R^2 d\Omega^2.\label{eq:Kruskal}
\end{equation}
On the hypersurface $U = 0$, the $UV$-component of the metric takes the form $\alpha^2 C_2 V^{-1/\mu}$, which is clearly finite and non-zero. 
Additionally, it is worth noting that for $\mu = 1/2$, we have $\gamma = 2$ and $\alpha = 1$. In this case, our localized coordinate system becomes identical to the standard Kruskal coordinates employed for an extremal Reissner-Nordstr{\"o}m spacetime~\cite{1966PhL....21..423C, Liberati:2000sq}.
Although the areal radius $R$ diverges as $U \to 0$, a particle moving radially toward $\rho = 0$ does not experience this divergence. Rather, this behavior implies that the particle's trajectory becomes purely radial when it reaches $\rho = 0$. This analysis shows that the geometry along the particle's path remains non-singular, with $U$ changing smoothly from a finite value to zero. Consequently, it is physically natural for the particle to traverse this boundary and enter the adjacent spacetime region where the sign of $U$ flips.
Given that particles can freely cross the extremal Killing horizon in the extremal RN geometry, it follows naturally that, at least for $\mu = 1/2$, an infalling particle will successfully traverse the $\rho = 0$ horizon.
To ensure a consistent regularization akin to the extremal RN geometry, this Kruskal-like coordinate extension requires selecting a specific analytic branch for the parameters $\alpha$ and $\gamma$, such that they satisfy
\begin{equation}
\begin{aligned}
    (-1)^\alpha &= -1,\\
    (-1)^\gamma &= 1.\label{eq:rag}
\end{aligned}
\end{equation}
It is straightforward to verify that the aforementioned branch conditions are identically satisfied when the parameter can be parameterized as $\mu = \frac{2n-1}{2m}$, where $m$ and $n$ are positive integers subject to the constraint $m > n$.
In terms of the parameter $c_{14}$, this algebraic requirement reads as
\begin{equation}
    c_{14} = 2- \frac{(2n-1)^2}{2m^2}.
\end{equation}

A similar Kruskal analysis applied to Eq.~\eqref{eq:gsol2} near $\rho = 0$ reproduces the same results of Eq.~\eqref{eq:Kruskal}. 
Therefore, whenever Eq.~\eqref{eq:rag} holds true, the extended geometry of Eq.~\eqref{eq:gsol} coincides precisely with Eq.~\eqref{eq:gsol2}, and the boundaries $U = 0$ and $V = 0$ exhibit a horizon-like structure reminiscent of the Schwarzschild black hole.
However, Eq.~\eqref{eq:rag} is only required for analytic continuation. If the objective is simply to smoothly glue the two spacetime manifolds together, this condition can be relaxed, provided that the explicit forms of the Kruskal metric \eqref{eq:Kruskal} are specified for different signs of $U$ and $V$.
Additionally, as $\rho \to \infty$, the tortoise coordinate $\rho^*$ under Eq.~\eqref{eq:gsol2} remains finite, indicating that the singularity at $\rho \to \infty$ is inherently space-like.

\begin{figure}[htb]
\centering

\begin{tikzpicture}[scale=2]

\draw[decorate, decoration={snake}] (-1,1)--(1,1);
\draw[decorate, decoration={snake}] (-1,-1)--(1,-1);

\draw[double] (-1,1)--(1,-1);
\draw[double] (-1,-1)--(1,1);

\draw (1,1)--(2,0);
\draw (-1,1)--(-2,0);
\draw (1,-1)--(2,0);
\draw (-1,-1)--(-2,0);

\draw[dashed] (1.0, -1.0)--
(0.845705, -0.787584)--
(0.72535, -0.606776)--
(0.634107, -0.450321)--
(0.568313, -0.311938)--
(0.525327, -0.186075)--
(0.503424, -0.0676797)--
(0.501725, 0.0479995)--
(0.520162, 0.165605)--
(0.559476, 0.289857)--
(0.621242, 0.425742)--
(0.707942, 0.578713)--
(0.823054, 0.754911)--
(0.971198, 0.961405);

\draw[dashed] (-1.0, 1.0)--
(-0.845705, 0.787584)--
(-0.72535, 0.606776)--
(-0.634107, 0.450321)--
(-0.568313, 0.311938)--
(-0.525327, 0.186075)--
(-0.503424, 0.0676797)--
(-0.501725, -0.0479995)--
(-0.520162, -0.165605)--
(-0.559476, -0.289857)--
(-0.621242, -0.425742)--
(-0.707942, -0.578713)--
(-0.823054, -0.754911)--
(-0.971198, -0.961405);

\node at (1,-1.1) {$i^-$};
\node at (1.6, -0.6) {$\mathscr{I}^-$};
\node at (2.1, 0.) {$i^0$};
\node at (1.6, 0.6) {$\mathscr{I}^+$};
\node at (1,1.2) {$i^+$};

\node at (0.4,0.6) {$S^+$};
\node at (0.4,-0.6) {$S^-$};
\node at (-0.2,-0.) {$S^0$};
\node at (0,1.2) {$S$};

\node at (0.9,-0.2) {Throat};

\end{tikzpicture}
\caption{Carter-Penrose diagram of the metric in Eq.~\eqref{eq:gsol} and its analytic extension in Eq.~\eqref{eq:gsol2} for $\mu < 1$. The double lines represent the shared boundary $\rho = 0$ (infinite physical radius), which acts as an extremal Killing horizon and escalates into a curvature singularity for $1/2 < \mu < 1$. The wavy lines $S$ denote the physical curvature singularities located at $r = 0$.}

\label{fig:Penrose}
\end{figure}
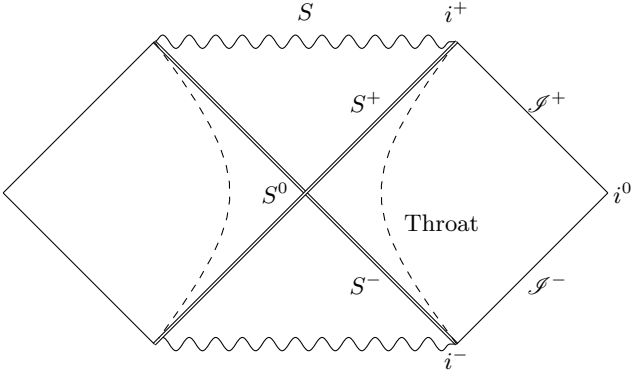

By performing this maximal analytic continuation, or alternatively, by smoothly gluing the two spacetime domains described by their respective metrics, we construct the comprehensive Carter-Penrose diagram illustrated in Fig.~\ref{fig:Penrose}.
Remarkably, while the resulting Carter-Penrose diagram closely resembles that of a standard Schwarzschild black hole, several profound distinctions emerge. First, a wormhole throat resides strictly outside the horizon. Second, the Killing horizon here corresponds to a spherical surface of infinite physical radius, and the surface gravity is identically zero. Furthermore, within the parameter range $1/2 < \mu < 1$, this horizon simultaneously escalates into a curvature singularity, situated at a finite proper distance from the wormhole throat; conversely, for $\mu \le 1/2$, the horizon becomes non-singular but is pushed to an infinite proper distance away from the throat. Crucially, under both parametric regimes, an infalling particle can cross this horizon within a finite proper time or affine parameter. Upon traversing the horizon, the signature inversion between the time and spatial coordinates forces the particle to inevitably culminate at the final spacelike singularity $S$.

\section{Discussion and Summary}\label{sec:summary}

In this work, we revisit the vacuum solutions of Einstein-Aether theory under a static aether configuration. 
Although this class of vacuum solutions has been explored in previous literature and is known to depend solely on the parameter $c_{14}$ as a one-parameter family, an explicit exact expression for arbitrary values of $c_{14}$, along with its complete analytic extension, has hitherto remained elusive.
By introducing a more suitable coordinate system, we successfully derived a remarkably concise and exact form for the vacuum solution as
\begin{equation}
\begin{aligned}
ds^2 =& -\rho^2 dt^2 + \frac{4 \mu ^4 R_s^2}{\rho ^4 \left(\rho ^{-\mu }-\rho ^{\mu }\right)^4} d\rho^2\\&
+\frac{\mu ^2 R_s^2}{\rho ^2 \left(\rho ^{-\mu }-\rho ^{\mu }\right)^2}(d\theta^2+\sin^2\theta d\phi^2),\label{eq:final1}
\end{aligned}
\end{equation}
where $\mu = \sqrt{1-c_{14}/2}$.
In the coordinates where the line element is given by $ds^2 = -g_{tt} dt^2 + g_{rr} dr^2 + r^2 d\Omega^2$, the metric components are parameterized by Eq.~\eqref{eq:metinr}.
Although the initial metric constantly maintains $g_{tt} > 0$, we obtain the corresponding branch with $g_{tt} < 0$ through analytic extension. The extended metric takes the form:
\begin{equation}
\begin{aligned}
ds^2 =& \rho^2 dt^2 - \frac{4 \mu ^4 R_s^2}{\rho ^4 \left(\rho ^{-\mu }+\rho ^{\mu }\right)^4} d\rho^2\\&
+\frac{\mu ^2 R_s^2}{\rho ^2 \left(\rho ^{-\mu }+\rho ^{\mu }\right)^2} 
(d\theta^2+\sin^2\theta d\phi^2),\label{eq:final2}
\end{aligned}
\end{equation}
and the metric components $g_{tt}$ and $g_rr$ are parameterized by Eq.~\eqref{eq:metinr2}.
The physical viability of these exact solutions strictly requires $\mu$ to be a real number, which translates into the parametric constraint $c_{14} \le 2$. Notably, the marginal case where $c_{14} = 2$ has already been comprehensively investigated in Ref.~\cite{Chan:2020amr}.

Utilizing this exact solution and the analytic extension, we systematically analyze the geometric properties of the spacetime. An essential conclusion read off from the metric is that the spacetime strictly rules out a black hole structure provided that $c_{14} \neq 0$.

For $c_{14} < 0$ ($\mu > 1$), the resulting asymptotically flat solution characterizes a naked singularity, with the characteristic profiles of $g_{tt}$ and $g_{rr}$ illustrated in Fig.~\ref{fig:2}. Within the sub-regime $c_{14} \le -6$ ($\mu \ge 2$), the spacetime hosts no photon sphere, rendering this naked singularity an SNS. Conversely, for $-6 < c_{14} < 0$, the naked singularity is securely enveloped by a photon sphere and is a WNS. Dynamic analysis reveals that a test particle will inevitably plunge into the naked singularity within a finite proper time or affine parameter; remarkably, this collision is also observed to occur within a finite coordinate time from the perspective of an asymptotic observer. 

For $c_{14} > 0$ ($\mu < 1$), the geometry exhibits a wormhole structure enveloped by a photon sphere, where the interior domain is non-asymptotically flat. 
The characteristic profiles of $g_{tt}$ and $g_{rr}$ is illustrated in Fig.~\ref{fig:1}.
For $0 < c_{14} < 3/2$ ($1/2 < \mu < 1$), the internal infinity acts as a curvature singularity at a finite proper distance from the throat. For $3/2 \le c_{14} < 2$ ($0 < \mu \le 1/2$), the internal infinity is regular but located at an infinite proper distance. However, as argued in Ref.~\cite{Eling:2006df}, this boundary remains a physical singularity due to the divergence of $R_{ab} k^a k^b$, with $k^a$ being the tangent vector of an affinely parameterized radial null geodesic approaching the internal infinity.
In both cases, test particles can arrive at the internal infinity in a finite proper time or affine parameter, while to an observer at external infinity, it takes an infinite amount of time.
Concurrently, this internal infinity functions as an extremal Killing horizon, characterized by a vanishing surface gravity.

We also discuss the geometry of the extended metric. For $c_{14} < 0$, the extended spacetime should be viewed as a cosmological-like solution that begins at a singularity and ends at another.
For the coupling regime $0 < c_{14} < 2$, the underlying geometry becomes considerably more intriguing. We find that in this scenario, the analytically extended metric can be smoothly glued together with the original metric across the extremal horizon, thereby constructing a global causal structure reminiscent of the maximally extended Schwarzschild black hole. In particular, when the coupling parameter satisfies the discrete quantized relation $c_{14} = 2 - \frac{(2n-1)^2}{2m^2}$, where $m$ and $n$ are positive integers subject to the constraint $m > n$, these two distinct spacetime domains act as the exact mutual analytic continuation of each other, precisely mirroring the global extension of an extremal Reissner-Nordström black hole. Crucially, a test particle moving along a radial geodesic experiences a perfectly regular spacetime metric in terms of the Kruskal coordinates; such a particle will successfully traverse this extremal Killing horizon and enter the newly uncovered, analytically extended spacetime region. The global causal architecture of this extension is illustrated by the Carter-Penrose diagram in Fig.~\ref{fig:Penrose}.

Furthermore, we must point out that the vacuum solutions derived in our present work are highly correlated with the classic Janis-Newman-Winicour (JNW) solution~\cite{Janis:1968zz, Virbhadra:1995iy}. To clarify this connection, we recall that the JNW vacuum solution represents the unique, static, and spherically symmetric spacetime geometry in GR minimally coupled to a massless real scalar field, with action
\begin{equation}
    S =  \int \sqrt{-g}\left(\frac{R}{16\pi G} - \frac{1}{2} \nabla_\mu \phi \nabla^\mu \phi\right). \label{eq:Saction}
\end{equation}
It can be directly verified that both metric configurations in Eq.~\eqref{eq:final1} and Eq.~\eqref{eq:final2} identically satisfy the equations of motion derived from this action, with the corresponding solution for the scalar field $\phi$ given by
\begin{equation}
    \phi(\rho) = \sqrt{\frac{\mu^2-1}{4\pi G}} \log \rho. \label{eq:Ssol}
\end{equation}
This means that when $\mu > 1$, the static AE vacuum geometry is identical to the JNW geometry. However, their physical meanings differ significantly: in AE theory, $\mu$ is a constant completely fixed by the coupling $c_{14}$, whereas in the JNW case, $\mu$ is a tunable parameter representing the scalar charge, making the JNW metric a two-parameter family of solutions.
Furthermore, for the JNW solution, the physical requirement of a real-valued scalar field rigorously constrains the parameter to the regime $\mu > 1$, under which the spacetime inevitably exhibits a naked singularity. Conversely, under the currently stringent experimental bounds $0 < c_{14} \lesssim 2.5 \times 10^{-5}$, the EA theory naturally selects the regime $\mu < 1$. This specific parametric domain unfolds a remarkably intricate global geometry, featuring a seamless transition from asymptotic flatness, through a wormhole-like throat, to an internal infinity that functions as an extreme Killing horizon, which eventually leads into a post-horizon space-like singularity. This paradigm shift precisely elucidates why such an exotic geometry remained unexplored until the present work, despite the JNW metric being discovered decades ago; the strict physical mandate for $\mu > 1$ within the canonical GR framework historically led researchers to dismiss the $\mu < 1$ domain as unphysical, thereby leaving its rich causal topology obscured.

In addition, when interpreting the metric configurations in Eq.~\eqref{eq:final2} and Eq.~\eqref{eq:Ssol} as an exact vacuum solution to the action in Eq.~\eqref{eq:Saction}, its underlying physical significance is remarkably profound: a massless scalar field minimally coupled to gravity is inherently capable of independently sustaining a cosmological-type geometry. This emergent spacetime originates from an initial singularity and terminates at a final singularity, with the scalar field intensity diverging identically at both singular boundaries.

Beyond the specific solution presented here, our analysis highlights several broader lessons. First, the present work demonstrates that an appropriate choice of coordinates can dramatically simplify the field equations and reveal previously hidden structures of the spacetime. In the static-aether sector of EA theory, a remarkably simple exact solution emerges only after adopting a suitable coordinate patch, making possible a complete analytic investigation of its global properties.
Second, our results indicate that the Schwarzschild black hole is an isolated configuration within this family of solutions. Once $c_{14}\neq0$, the spacetime no longer describes a black hole geometry. Instead, depending on the sign of $c_{14}$, the solution either becomes a naked singularity or develops a wormhole-like structure whose throat lies outside the horizon. This suggests that even a small departure from the GR limit qualitatively changes the global nature of the vacuum spacetime.
Finally, by constructing the analytic extension of the solution, we uncover its complete spacetime structure. For $0<c_{14}<2$, the asymptotically flat exterior region is connected through a wormhole throat to an internal infinity that acts as an extremal Killing horizon. Beyond this horizon lies a new spacetime domain in which the causal roles of temporal and radial directions are interchanged, ultimately terminating at a final spacelike singularity. To our knowledge, this provides an explicit example where an observer can traverse an extremal Killing horizon and enter a region with a reversed causal character, thereby revealing a much richer global structure than previously recognized.

\section*{Acknowledgements}

We are grateful to Lirui Yang for inspiring discussions.
This work was supported in part by the National Natural Science Foundation of China under Grant No.~12547101. HL was also supported by the start-up fund of Chongqing University under No.~0233005203009, and JZ was supported by the start-up fund of Chongqing University under No.~0233005203006.

\appendix

\bibliography{refs}

\end{document}